\documentclass[groupaddress,pra]{revtex4-1}

\reversemarginpar

\newcommand{\be}{\begin{equation}}
\newcommand{\ee}{\end{equation}}
\newcommand{\benn}{\begin{displaymath}}
\newcommand{\eenn}{\end{displaymath}}

\usepackage{graphicx,tabularx}
\usepackage{xspace}
\usepackage{dcolumn}
\usepackage{bm}
\usepackage[version=3]{mhchem}
\usepackage{verbatim}
\usepackage[colorlinks=true, urlcolor=blue, citecolor=blue, filecolor=blue, linkcolor=blue]{hyperref}
\usepackage{amsmath}
\usepackage{amssymb}
\usepackage{multirow}
\usepackage{natbib}
\usepackage[none]{hyphenat}
\usepackage{threeparttable}
\usepackage{color}

\def\sc {\scriptscriptstyle}
\def\spc {\hskip  0.25mm}
\def\spm {\hskip -0.25mm}

\def\Q {\mathcal{Q}}

\def\O {\mathit \Omega}
\def\v {\boldsymbol{v}}
\def\s {\boldsymbol{\sigma}}

\def\J {\boldsymbol{\mathcal{J}}}
\def\divv {\nabla\!\spm\spm \cdot\spm\spm \boldsymbol{v}}

\def\devv {\substack{\sc{\circ} \\[-0.2mm] \displaystyle \overline{\nabla \spm \boldsymbol{v}}\\[6pt] }}

\def\g {\textsl{g}}

\begin{document}

\title{Rayleigh-Brillouin scattering in binary mixtures of disparate-mass constituents: SF$_6-$He, SF$_6-$D$_2$ and SF$_6-$H$_2$}

\author{Yuanqing Wang, Wim Ubachs}
\address{Department of Physics and Astronomy, LaserLaB, Vrije Universiteit, De Boelelaan 1081, 1081 HV Amsterdam, The Netherlands}

\author{Cesar A. M. de Moraes, Wilson Marques Jr}
\address{Departamento de F\'{i}sica, Universidade Federal do Paran\'{a}, Caixa Postal 19044, Curitiba 81531-990, Brazil}

\date{\today}

\begin{abstract}

\noindent The spectral distribution of light scattered by microscopic thermal fluctuations in binary mixture gases was investigated experimentally and theoretically. Measurements of Rayleigh-Brillouin spectral profiles were performed at a wavelength of 532 nm and at room temperature, for mixtures of SF$_6-$He, SF$_6-$D$_2$ and SF$_6-$H$_2$.
In these measurements, the pressure of the gases with heavy molecular mass (SF$_6$) is set at 1 bar, while the pressure of the lighter collision partner was varied.
In view of the large polarizability of SF$_6$ and the very small polarizabilities of He, H$_2$ and D$_2$, under the chosen pressure conditions these low mass species act as spectators and do not contribute to the light scattering spectrum, while they influence the motion and relaxation of the heavy SF$_6$ molecules.
A generalized hydrodynamic model was developed that should be applicable for the particular case of molecules with heavy and light disparate masses, as is the case for the heavy SF$_6$ molecule, and the lighter collision partners. Based on the kinetic theory of gases, our model replaces the classical Navier-Stokes-Fourier relations with constitutive equations having an exponential memory kernel. The energy exchange between translational and internal modes of motion is included and quantified with a single parameter $z$ that characterizes the ratio between the mean elastic and inelastic molecular collision frequencies. The model is compared with the experimental Rayleigh-Brillouin scattering data, where the value of the parameter $z$ is determined in a least-squares procedure. Where very good agreement is found between experiment and the generalized hydrodynamic model, the computations in the framework of classical hydrodynamics strongly deviate. Only in
the hydrodynamic regime both models are shown to converge.

\end{abstract}


\maketitle

\section{Introduction}

Spontaneous Rayleigh-Brillouin (RB) scattering spectra of gaseous/fluidic media contain information on the thermodynamics and the transport properties of the gases/fluids such as thermal diffusivity, speed of sound, relaxation times of various dynamical processes, etc.~\cite{Miles2001,Mcmanus2019,Bruno2019}. In addition, such spectra provide a way to test a given theory of the energy relaxation dynamics in gases, and in gaseous mixtures.
Theories of RB-scattering for fluids in the kinetic and hydrodynamic regimes have been devised based on the linearized Boltzmann equation and on linearized hydrodynamic equations, respectively. Many models have already successfully explained the scattered light spectra~\cite{Yip1964,Mountain1966,Marques1998}.
Some models have accomplished to reproduce the spectra very well, even though some discrepancies with experiments still exist, while their validity is often restricted to certain pressure regimes.

A celebrated example of a model representation of RB-scattering profiles in the kinetic regime is the Tenti-model~\cite{Boley1972,Tenti1974}, which was applied to describe the spectral profiles of a variety of gases, such as N$_2$~\cite{GU2013b}, CO$_2$~\cite{Gu2014a,Wang2019}, and N$_2$O~\cite{Wang2018}. However, the Tenti-model was not developed for modeling the RB-spectra of mixture gases. Moreover, it requires as input the values of the macroscopic transport coefficients, such as heat capacity, thermal conductivity, specific heat ratio, and shear viscosity, in general not known for composite mixtures. And ultimately the value of the bulk viscosity, the parameter determining internal relaxation, is typically determined via a fitting procedure when applying the Tenti-model. This makes that the Tenti-model is not immediately applicable to the present case of binary mixtures.

The hydrodynamic theory of light scattering in binary fluid mixtures was first developed by~\citet{Mountain1969}, who described the local dielectric constant fluctuations by several linear hydrodynamic equations including the continuity equation for mass conservation, the Navier-Stokes equation for momentum conservation, the diffusion equation and the energy transport equation. Later, \citet{Cohen1971} corrected some correlation functions by adding the 'non-Lorentzian' term based on the original paper of~\citet{Mountain1969} therewith improving the light scattering models.

On the experimental side Rayleigh-Brillouin scattering spectra of helium-xenon atomic gas mixtures were measured by \citet{Letamendia1981} at different pressures, compositions, and scattering angles. The data were compared with a complete two-component hydrodynamic theory and good agreement was found at low molar fractions of He and at molar fractions of He higher than a 'critical' value which depends on the partial pressure of Xe. In addition, since the spectral shape in He$-$Xe mixtures is very sensitive to the presence and magnitude of thermal-diffusion effects, the thermal diffusion coefficient could be derived.
A kinetic model was formulated by \citet{Letamendia1985} based on the generalized Enskog equations for a binary mixture of hard-sphere fluids. This model gives an improvement over an existing model derived by \citet{Boley1972b}, which is based on the linearized Boltzmann equations for Maxwell molecules and which was successful in explaining light scattering spectra of He$-$Xe mixtures at very low Xe pressure and small Xe molar fraction, conditions under which imperfect-gas effects and thermal diffusion can be ignored.

\citet{Bonatto2005} proposed a model to describe the spontaneous density fluctuations in a binary mixture of monatomic ideal gases based on the Boltzmann equation, the collision operators of which are replaced by simple relaxation-time terms. For this model, the description of kinetic equations for a mixture of monatomic ideal gases is characterized by the fields of partial number density, partial flow velocity and partial temperature and assuming that the particles of the mixture interact according to the Lennard-Jones (6$-$12) potential. This model was applied to the light scattering spectrum of a binary gas mixture passing over from a hydrodynamic to a kinetic regime. The measurements by \citet{Gu2015} of RB-spectra on mixtures of Ar$-$He and Kr$-$He, were found to produce excellent agreement with this model. Clearly, since noble gas atoms do not have the internal degrees of motion that molecules have, this model is not suited for mixtures including molecular gases.

In order to extend these studies into the molecular regime by including intra-molecular as well as inter-molecular relaxation, RB-scattering of mixtures of SF$_6-$He, SF$_6-$D$_2$, SF$_6-$H$_2$ is measured under different conditions. A relaxation hydrodynamic model for these mixtures of specific disparate masses is developed, based on a generalized hydrodynamic description, and a comparison will be made between the experimental data and the model developed.
Also a comparison will be made with a classical hydrodynamics model.

\section{Relaxation hydrodynamic model for binary mixtures}
\label{sec:MixtureModel}

In a fluid in thermal equilibrium, the intensity of the Rayleigh-Brillouin scattered light is related to the fluctuations of the dielectric constant $\delta \epsilon$ caused by the random
thermal motion of molecules~\cite{Fabelinskii2012}. For a binary gas mixture these fluctuations are related to the fluctuations of thermodynamic variables as pressure $p$, temperature $T$ and mass concentration of one constituent $c$:
\begin{equation}\label{Eq:DielectriFlucMixture}
  \delta \epsilon(\boldsymbol r, t) = \left(\frac{\partial \epsilon}{\partial p}\right)_{\!\!\spm {\sc T},c}\delta p(\boldsymbol r, t) + \left(\frac{\partial \epsilon}{\partial T}\right)_{\!\!\spm p,c}\delta T(\boldsymbol r, t) + \left(\frac{\partial \epsilon}{\partial c}\right)_{\!\!\spm p,{\sc T}}\delta c(\boldsymbol r, t)
\end{equation}
and the dynamic structure factor provides an expression from which the scattering spectrum can be computed:
\begin{equation}\label{Eq:StructureMixture}
  S(\boldsymbol q, \omega) = 2\spc {\rm Re}\spc [ \langle \delta  \epsilon(\boldsymbol q, i\omega) \delta \epsilon(- \boldsymbol q, 0) \rangle ]
\end{equation}
where $\boldsymbol q$ represents the scattering vector with magnitude:
\begin{equation}
  q = 2k_i \sin{\frac{\theta}{2}} = \frac{4\pi n}{\lambda_i} \sin{\frac{\theta}{2}}
  \label{q-vec}
\end{equation}
with $k_i$ and  $\lambda_i$ the wave vector and wavelength of the incident light, $n$ the refractive index and $\theta$ the scattering angle~\cite{Wang2020}. This sets the dependence of the dynamic structure factor on the experimental conditions.

In classical mixture theory \cite{Mountain1969,Cohen1971}, the macroscopic state of a binary gas mixture is characterized by the six scalar fields of mass density $\rho=\rho_{1}+\rho_{2}$, flow velocity $\boldsymbol v$ (contributing a field for each direction), temperature $T$ and mass concentration $c=\rho_{1}/\rho$, where the index 1 refers to light constituent, while the index 2 refers to the heavy one. The balance equations governing the dynamical behavior of these fields are:\\
(1) the continuity equation:
\begin{equation}\label{Eq:Continuity}
  \frac{{\rm d} \rho}{{\rm d} t} + \rho \nabla \cdot \boldsymbol v = 0,
\end{equation}\\
(2) the momentum equation:
\begin{equation}\label{Eq:NavierStokes}
  \rho \frac{{\rm d}\boldsymbol v}{{\rm d} t} + \nabla \cdot \boldsymbol \sigma  = 0,
\end{equation}\\
(3) the diffusion equation:
\begin{equation}\label{Eq:Diffusion}
  \rho \frac{{\rm d}c}{{\rm d} t} + \nabla \cdot \J = 0,
\end{equation}\\
(4) the energy transport equation:
\begin{equation}\label{Eq:EnergyTrans}
 \rho \frac{{\rm d}\varepsilon}{{\rm d} t} + \nabla  \cdot \boldsymbol  \kappa +  \boldsymbol \sigma:\nabla \boldsymbol{v} = 0,
\end{equation}\\
where $\boldsymbol{\sigma}$ is the pressure tensor, $\varepsilon$ is the mixture specific internal energy, $\boldsymbol{\kappa}$ is the heat flux vector and $\J$ is the diffusion flux of the light constituent in the mixture, while ${\rm d}/{\rm d}t=\partial/\partial t+\v\cdot \nabla$ denotes the material time derivative.
The balance equations~\eqref{Eq:Continuity}-\eqref{Eq:EnergyTrans} become a closed set of field equations for the determination of the basic fields if we provide constitutive relations for the pressure tensor, the heat flux vector and the diffusion flux.
In the Navier-Stokes-Fourier approximation, these constitutive relations are~\cite{deGrootMazur1969}:\\
(i) the Navier-Stokes law
\begin{equation}\label{Eq:PressureTensor}
  \boldsymbol \sigma = (p - \eta_{\rm b} \nabla \cdot \boldsymbol v)\boldsymbol I - 2\ \eta_{\rm s}\devv,
\end{equation}
where $p$ is the mixture pressure, $\eta_{\rm b}$ is the volume (or bulk) viscosity, $\eta_{\rm s}$ is the shear viscosity and $\devv$ is rate-of-shear tensor;\\
(ii) the Fourier law of heat conduction:
\begin{equation}\label{Eq:HeatFlux}
\boldsymbol \kappa = -\lambda_{\sc\rm th} \nabla T + \left[\mu-T\left(\frac{\partial \mu}{\partial T}\right)_{p,c}+ k_{\sc T}\left(\frac{\partial \mu}{\partial c}\right)_{p,\sc T} \right] \J,
\end{equation}
where $\lambda_{\rm th}$ is the thermal conductivity of the mixture, $k_{\sc T}$ is the thermal diffusion ratio \cite{Landau1987a} and
$\mu=\mu_{\sc 1}/m_{\sc 1}-\mu_{\sc 2}/m_{\sc 2}$ is the mixture chemical potential (i.e., the difference in the chemical potential per
unit mass of the two constituents);  \\
(iii) the Fick law of diffusion
\begin{equation}\label{Eq:DiffusionFluxFick}
  \J = -\rho D_{\sc 12}\left[ \nabla c + \frac{k_{p}}{p} \nabla p+\frac{k_{\sc T}}{T} \nabla T\right],
\end{equation}
where $D_{\sc 12}$ is the diffusion coefficient and
$k_{p}\!=\!-(p/\spm\spm \rho^{\spc \sc 2})(\partial\rho/\partial c)_{\sc p,T}/(\partial\mu/\partial c)_{\sc p,T}$ is the so-called barodiffusion factor.

As pointed out in the literature~\cite{Weiss1995}, the description of time-dependent processes based on the Navier-Stokes-Fourier theory starts to deviate from experimental data at high frequencies, such as in the case of Rayleigh-Brillouin scattering, where typically hypersound frequencies come into play. In order to overcome this situation, we may consider a generalization of the Navier-Stokes-Fourier constitutive relations by assuming that the pressure tensor, the heat flux vector and the diffusion flux respond to gradients only after a relaxation time has elapsed. Such type of approach was first introduced by \citet{Cattaneo1958} to address the paradox of heat conduction in single fluid which follows from Fourier's law and it leads to a constitutive equation for the heat flux vector with an exponential memory kernel. In the present paper, our generalized constitutive equations follow from the kinetic theory proposed by \citet{Alievskii1969} for a mixture of polyatomic gases. Basing on Grad’s moment method~\cite{Grad1949}, a macroscopic state of a mixture is characterized by the basic fields of partial mass densities, partial diffusion fluxes, partial stress tensors, partial specific energies and partial heat fluxes associated with translational and internal molecular degrees of freedom.

A closed system of linear field equations for the determination of the basic fields can be obtained if we multiply the Boltzmann equation for a polyatomic gas mixture by an appropriate set of Hermite and internal energy polynomials, integrate over peculiar velocities and sum over internal states.
The collision integrals appearing in these equations can be
expressed in terms of the basic fields by considering the so-called 17-moment approximation to the distribution function. In the case of a binary gas mixture, where the molecular mass ratio $m_{\sc 1}/m_{\sc 2}$ is a small parameter and the exchange of energy between translational and internal degrees of freedom of the molecules is a slow process, this system of field equations becomes fully decoupled and its solution can be used to derive the following constitutive relations:
\begin{gather}
\label{Eq:ExpMemoryPressureTensor}
\s=(\spc p-\!\!\int_{\spm \sc 0}^{\spc t} \!\!\eta_{\rm b} (t-t^\prime)\spc \divv (\boldsymbol{r},t^\prime)\spc dt^\prime\spc )\spc \boldsymbol{I}
-2\spc [\eta_{\rm s}]_{_{\sc 1}}\devv
-2\spc \int_{\spm \sc 0}^{\spc t} \!\![\eta_{\rm s}]_{_{\sc 2}} (t-t^\prime)\spc \devv (\boldsymbol{r},t^\prime)\spc dt^\prime,
\end{gather}
\begin{gather}
\label{Eq:ExpMemoryHeatFlux}
\boldsymbol \kappa=-[\lambda_{\rm th}]_{_{\sc 1}} \spm \nabla T
-\int_{\spm \sc 0}^{\spc t}\! [\lambda_{\rm th}]_{_{\sc 2}} (t-t^\prime) \nabla T(\boldsymbol{r},t^\prime)\spc dt^\prime
+\Bigl[\mu-T \left (\frac{\partial\mu}{\partial T}\right)_{p,c}+k_{\sc T} \left(\frac{\partial \mu}{\partial c}\right)_{\sc p,T}\Bigr]\spc \J,
\end{gather}
\begin{gather}
\label{Eq:ExpMemoryDiffusionFluxFick}
\J=-\rho\!\int_{\spm \sc 0}^{\spc t}\! D_{\spm \sc 12} (t-t^\prime)
\Bigl[ \nabla\spm c\spc (\boldsymbol{r},t^\prime) +\frac{k_{p}}{p}\spc \nabla\spm p\spc (\boldsymbol{r},t^\prime)
+\frac{k_{\sc T}}{T}\spc \nabla T(\boldsymbol{r},t^\prime)\Bigr]\spc  dt^\prime,
\end{gather}
where the generalized transport coefficients are defined as
\begin{gather}
\eta_{\rm b} (t-t')=\eta_{\rm b} \spc
\frac{ e^{\displaystyle -(t-t')/\tau_{\spm \sc v}}}{\tau_{\spm \sc v}},
\end{gather}

\begin{gather}
[\eta_{\rm s}]_{_{\sc 2}} (t-t')=[\eta_{\rm s}]_{_{\sc 2}}\spc \frac{e^{\displaystyle -(t-t')/\tau_{\spm \sc 2}}}{\tau_{\spm \sc 2}},
\end{gather}

\begin{gather}
[\lambda_{\sc\rm th}]_{_{\sc 2}} (t-t')=
[\lambda_{\sc\rm th}^{\sc\rm {tr}}]_{_{\sc 2}} \spc
\frac{e^{\displaystyle-(t-t')/\tau_{\spm \sc 2}^\prime}}{\tau_{\spm \sc 2}^\prime}
+[\lambda_{\sc\rm th}^{\sc\rm {int}}]_{_{\sc 2}} \spc
\frac{e^{\displaystyle -(t-t')/\tau_{\spm \sc 2}^{\prime\prime}}}{\tau_{\spm \sc 2}^{\prime\prime}},
\end{gather}

\begin{gather}
D_{\sc 12}(t-t')=D_{\sc 12}\spc
\frac{e^{\displaystyle -(t-t')/\tau_{\spm \sc w}}}{\tau_{\spm \sc w}}.
\end{gather}
It is clear that our generalized constitutive relations depend on exponential memory kernels which are connected with the characteristic relaxation times $\tau_{\spm \sc v}$, $\tau_{\spm \sc 2}$, $\tau_{\spm \sc 2}^\prime$, $\tau_{\spm \sc 2}^{\prime\prime}$ and $\tau_{\spm \sc w}$ that give us, respectively, a measure of time interval spent by dynamic pressure, partial stress tensors, partial heat fluxes and diffusion flux to achieve a stationary value.

Insertion of the constitutive relations~\eqref{Eq:ExpMemoryPressureTensor}-\eqref{Eq:ExpMemoryDiffusionFluxFick} into the conservation equations~\eqref{Eq:Continuity}-\eqref{Eq:EnergyTrans} leads to a linear system of field equations. As ($p$, $c$, $T$) is not a set of statistically independent variables, here it is replaced by ($\phi$, $p$, $c$) with:
\begin{equation*}
  \phi = T - \frac{T_{\sc 0} \beta_{\sc T}}{\rho_{\sc 0} c_p}p
\end{equation*}
where $\beta_{\sc T}=-\rho_{\sc 0}^{ -1} (\partial \rho/\partial T)_{\sc p,c}= T_{\sc 0}^{-1}$ and $c_p=(\partial \varepsilon/\partial T)_{\rho,c}+T_{\sc 0}\spc (\partial p/\partial T)_{\!\rho,c}^{2}
/\rho_{\sc 0}^{2}(\partial p/\partial \rho)_{{\sc T},c}$ are the thermal expansion coefficient and the specific heat capacity at constant pressure, respectively. Equilibrium values are denoted by the subscript zero, while thermodynamics derivatives are understood to be evaluated at equilibrium. After Fourier-Laplace transformations, this linear system for macroscopic fluctuations can be rewritten as:
\begin{equation}\label{Eq:FLtransform}
  \boldsymbol{ A  \psi}(\boldsymbol q, s) = \boldsymbol{ B  \psi}(\boldsymbol q, 0)
\end{equation}
where
\begin{gather}
\boldsymbol{\psi}=
\begin{pmatrix}
\psi_{\sc 1}\\[4mm] \psi_{\sc 2}\\[4mm] \psi_{\sc 3}\\[4mm] \psi_{\sc 4}
\end{pmatrix}=
\begin{pmatrix}
\beta_{\sc T}\spc \bar{\phi}\\[4mm] \bar{c}\\[4mm] \bar{p}/p_{\sc 0}\\[4mm] \tau_{\spm\spm \sc s}\nabla\!\spm\spm \cdot\spm\spm \bar{\boldsymbol{v}}
\end{pmatrix}
\end{gather}
and the $4\spm\times\spm 4$ matrices $\boldsymbol{A}$ and $\boldsymbol{B}$ have the form
\begin{equation}\label{Eq:MatrixA}
  \boldsymbol A = \left(
  \begin{array}{cccc}
  s + f(s) q^2& -s  \dfrac{\gamma-1}{\gamma} \rho_{\sc 0} \dfrac{k_{\sc T}}{p_{\sc 0}} \left(\dfrac{\partial \mu}{\partial c}\right)_{\!\!\spm p,\sc T}& \dfrac{\gamma-1}{\gamma} f(s) q^2 &0 \\[2mm]
  k_{\sc T} \g(s) q^2&s + \g(s) q^2& \mathcal{P} k_{p} \g(s) q^2& 0\\[2mm]
  -s \gamma &-s \gamma \rho_{\sc 0} \dfrac{k_{p}}{p_{\sc 0}} \left(\dfrac{\partial \mu}{\partial c}\right)_{\!\!\spm p,\sc T} &s& \gamma\\[2mm]
  0&0&-q^2& s+b(s)q^2
  \end{array}
  \right)
\end{equation}
\begin{equation}\label{Eq:MatrixB}
  \boldsymbol B = \tau_s
  \begin{pmatrix}
    1& -\dfrac{(\gamma-1)}{\gamma}\rho_{\sc 0} \dfrac{k_{\sc T}}{p_{\sc 0}} \left(\dfrac{\partial \mu}{\partial c}\right)_{\!\!\spm p,\sc T}&0&0 \\[2mm]
   0 &1&0& 0\\[2mm]
  -\gamma & -\gamma \rho_{\sc 0}\dfrac{k_p}{p_{\sc 0}} \left(\dfrac{\partial \mu}{\partial c}\right)_{\!\!\spm p,\sc T}&1&0 \\[2mm]
  0&0&0&1
  \end{pmatrix}
\end{equation}
In the above expressions, time is given in units of the stress relaxation time:
\begin{equation}
    \tau_{\spm s}=(\spc [\eta_{\rm s}]_{_{\sc 1}}+[\eta_{\rm s}]_{_{\sc 2}})/p=\eta_{\rm s}/p
    \label{eq-tau_s}
\end{equation}
and length is given in units of the mixture mean free path $\tau_{s}\sqrt{p_{\sc 0}/\spm\rho_{\sc 0}}$. The specific heat capacity ratio of the mixture:
\begin{equation}
\label{gammix}
 \gamma=1+[\spc x_{\sc 1}/(\gamma_{\sc 1}-1)+x_{\sc 2}/(\gamma_{\sc 2}-1)\spc ]^{-1}
\end{equation}
can be calculated from the specific heat ratios $\gamma_{\sc i}$ (as listed in Table~\ref{Tab:PolarizabilityandGamma}) and the molar fractions $x_{\sc i}$ of the constituents. In the above matrices the following functional forms are represented:
\begin{equation}
  \mathcal{P} = 1+ \frac{\gamma -1}{\gamma}\frac{k_{\sc T}}{k_{p}}
\end{equation}
\begin{equation}
  \g(s) = \frac{\rho_{\sc 0} D_{\sc 12}}{\eta_{\rm s}}\frac{\tau_{\spm s}/\tau_{\spm \sc w}}{s+\tau_{\spm s}/\tau_{\spm \sc w}}
  \label{g-rhm}
\end{equation}
\begin{equation}
  f(s) = \frac{[\lambda_{\rm th}]_{_{\sc 1}}}{\eta_{\rm s}c_p} +
  \frac{[\lambda_{\rm th}^{\rm tr}]_{_{\sc 2}}}{\eta_{\rm s}c_p}\frac{\tau_{\spm s}/\tau_{\sc 2}^\prime}{s+\tau_{\spm s}/\tau_{\sc 2}^\prime}+
  \frac{[\lambda_{\rm th}^{\rm int}]_{_{\sc 2}}}{\eta_{\rm s}c_p}\frac{\tau_{\spm s}/\tau_{\sc 2}^{\prime\prime}}{s+\tau_{\spm s}/\tau_{\sc 2}^{\prime\prime}}
  \label{f-rhm}
\end{equation}
\begin{equation}
  b(s) = \frac{4}{3}\frac{[\eta_{\rm s}]_{_{\sc 1}}}{\eta_{\rm s}} +  \frac{4}{3}\frac{[\eta_{\rm s}]_{_{\sc 2}}}{\eta_{\rm s}}\frac{\tau_{\spm s}/\tau_{\sc 2}}{s+\tau_{\spm s}/\tau_{\sc 2}} + \frac{\eta_{\rm b}}{\eta_{\rm s}}\frac{\tau_s/\tau_{\sc v}}{s+\tau_s/\tau_{\sc v}}
  \label{b-rhm}
\end{equation}
Note that the light scattering predictions of the classical mixture theory derived by Mountain and Deutch~\cite{Mountain1969} follows from our generalized hydrodynamic model if we set the functions appearing in matrix
$\boldsymbol{A}$ as
\begin{equation}
    \g(s) = \frac{\rho_{\sc 0} D_{\sc 12}}{\eta_{\rm s}}
    \label{g-clas}
\end{equation}
\begin{equation}
    f(s) = \frac{[\lambda_{\rm th}]_{_{\sc 1}}+
    [\lambda_{\rm th}^{\rm tr}]_{_{\sc 2}} + [\lambda_{\rm th}^{\rm int}]_{_{\sc 2}}}{\eta_{\rm s}c_p}
    \label{f-clas}
\end{equation}
\begin{equation}
  b(s) = \frac{4}{3} + \frac{\eta_{\rm b}}{\eta_{\rm s}}
  \label{b-clas}
\end{equation}

For the computation of the RB-spectral profiles the matrix equation can be further evaluated. This can be done, both for the relaxation hydrodynamics model taking the functional forms for $\g(s)$, $f(s)$  and $b(s)$ as defined in Eqs.~\eqref{g-rhm}-\eqref{b-rhm},
as well as for the classical hydrodynamics model, by taking the forms defined in Eqs.~\eqref{g-clas}-\eqref{b-clas}. Here we proceed by evaluating the more complex case for the relaxation hydrodynamics model.
Since the solution of Eq.~\eqref{Eq:FLtransform} can be cast in the form:
\begin{equation}
\label{Eq:SoutionofFLtransform}
  \psi_i(\boldsymbol q, s) = \sum_r\Q_{ir}\psi_r(\boldsymbol q, 0),
\end{equation}
where $\boldsymbol \Q = \boldsymbol A^{-1}\boldsymbol B$,
the correlation functions of the form $\langle \psi_i({\boldsymbol q}, s)\psi_j(-{\boldsymbol q}, 0)\rangle$ follow as:
\begin{equation}\label{Eq:CorrelationFunc}
  \langle \psi_i(\boldsymbol q, s)\psi_j(-\boldsymbol q, 0)\rangle =
  \sum_r \Q_{ir}\langle \spc \psi_r(\boldsymbol q, 0) \psi_j (-\boldsymbol{q},0)\spc \rangle
\end{equation}
where the equal-time correlation functions, which follow from the thermodynamic theory of fluctuations, read:
\begin{eqnarray}
       \langle \spc |\psi_{\sc 1}(\boldsymbol q, 0)|^2 \rangle &=& \frac{V^2}{N_{\sc 0}} \frac{(\gamma-1)}{\gamma}\\
        \langle |\psi_{\sc 2}(\boldsymbol q, 0)|^2 \rangle &=& \frac{V^2}{N_{\sc 0}}
        x_{\sc 1} x_{\sc 2} \frac{(m_{\sc 1} m_{\sc 2})^2}{(m_{\sc 1} x_{\sc 1}
        +m_{\sc 2} x_{\sc 2})^4}\\
        \langle \spc |\psi_{\sc 3} (\boldsymbol q, 0)|^2 \rangle &=&
        \frac{V^2}{N_{\sc 0}} \gamma
\end{eqnarray}
where $N_{\sc 0}$ is the total number of molecules in the volume $V$ of the scattering region. In terms of the set of dimensionless variables ($\psi_{\sc 1}$, $\psi_{\sc 2}$, $\psi_{\sc 3}$) we can now write the dynamic structure factor of~Eq.~\eqref{Eq:StructureMixture} as:

\begin{equation}\label{Eq:FinalStructureMixture}
   S(\boldsymbol q, \omega) = 2 \sum_{ij} \left( \frac{\partial \epsilon}
   {\partial \psi_i}\right)  \left( \frac{\partial \epsilon}{\partial \psi_j}\right) \langle |\psi_j(\boldsymbol q, 0)|^2 \rangle {\rm Re}\spc [\Q_{ij}(s=i\omega)].
\end{equation}
Note that the matrix element containing $\psi_4$ needs not be evaluated since it does not appear in the structure factor  $S(\boldsymbol q, \omega)$. For a binary mixture obeying the Clausius-Mossotti relation (see for example the textbook of Born~\cite{Born2013}) we can obtain the following relations:
\begin{eqnarray}
       \left(\frac{\partial \epsilon}{\partial \psi_{\sc 1}}\right) &=& -\frac{N_{\sc 0}}{V}\spc (\alpha_{\sc  1} x_{\sc 1} + \alpha_{\sc  2} x_{\sc 2}) \\
       \left(\frac{\partial \epsilon}{\partial \psi_{\sc 2}}\right) &=& \frac{N_{\sc 0}}{V}\spc \frac{(m_{\sc 1} x_{\sc 1}+m_{\sc 2} x_{\sc 2})^2}{m_{\sc 1} m_{\sc 2}}\spc (\alpha_{\sc  1}- \alpha_{\sc  2}) \\
       \left(\frac{\partial \epsilon}{\partial \psi_{\sc 3}}\right)&=&\frac{N_{\sc 0}}{V}\spc \frac{1}{\gamma}\spc (\alpha_{\sc  1} x_{\sc 1} + \alpha_{\sc  2} x_{\sc 2})
     \end{eqnarray}
in which $\alpha_{\sc  1}$, $\alpha_{\sc  2}$ are the dynamic polarizabilities of the two molecular species at the frequency of the incident light. The molecular polarizabilities at 532.22 nm can be found in Table~\ref{Tab:PolarizabilityandGamma}.

\begin{table}
  \centering
  \renewcommand\tabcolsep{20.0pt}
  \caption{The dynamic polarizabilities $\alpha$ (given in units of $10^{-40}$ C$\cdot$ m$^2$/V) at wavelength of 532.22 nm were calculated based on the static polarizability and the dynamic polarizability function~\cite{Lide2004}, and the used values as cited in footnotes.
  Heat capacity ratio $\gamma$ for the molecular species as obtained from experiment.}\label{Tab:PolarizabilityandGamma}
  \begin{threeparttable}
  \begin{tabular}{c c c}
  \hline
  molecule & $\alpha$  & $\gamma$ \\
  \hline
  SF$_6$& 5.029\tnote{a}  & 1.10\tnote{d}\\
  He    & 0.231\tnote{b}  & 1.66\tnote{e} \\
  D$_2$ & 0.900\tnote{c}  & 1.40\tnote{e}\\
  H$_2$ & 0.911\tnote{c}  & 1.41\tnote{e}\\
  \hline
  \end{tabular}
  \begin{tablenotes}
        \footnotesize
        \item[a] Ref.~\cite{Bridge1964}.
        \item[b] Ref.~\cite{Chung1968}.
        \item[c] Ref.~\cite{Bridge1997}.
        \item[d] Ref.~\cite{Yokomizu2015}.
        \item[e] Ref.~\cite{Koehler1950}.
    \end{tablenotes}
\end{threeparttable}
\end{table}

This model is tightly related to the transport coefficients of the binary mixture: the bulk viscosity $\eta_{\rm b}$, the shear viscosity $\eta_{\rm s}= [\eta_{\rm s}]_{_{\sc 1}}+[\eta_{\rm s}]_{_{\sc 2}}$, the thermal conductivity $\lambda_{\rm th}$, the diffusion coefficient $D_{\spc \sc 12}$ and the thermal diffusion ratio $k_T$.
The characteristic relaxation times and the usual transport coefficients appearing in the generalized constitutive relations of Eqs.~\eqref{Eq:ExpMemoryPressureTensor}-\eqref{Eq:ExpMemoryDiffusionFluxFick} are given by:
\begin{gather}
\label{tau-v}
\eta_{\rm b}=\frac{3}{2}\spc (\gamma-1)\spc (5/3-\gamma)\spc p \spc \tau_{\spm\spm \sc v},
\end{gather}
\begin{gather}
\label{eta-s1}
[\eta_{\rm s}]_{_{\sc 1}}=\frac{5}{8}\spc \frac{x_{\sc 1} k_{\spm \sc \rm B} T}{x_{\sc 1} \O_{\sc 11}^{\sc (2,2)}+2\spc x_{\sc 2} \O_{\sc 12}^{\sc (2,2)}}
\end{gather}
\begin{gather}
[\eta_{\rm s}]_{_{\sc 2}}=p\spc x_{\sc 2} \tau_{\sc 2}=\frac{5}{8}\spc \frac{k_{\spm \sc \rm B} T}{\O_{\sc 22}^{\sc (2,2)}},
\end{gather}

\begin{gather}
D_{\spm \sc 12}=k_{\spm \sc \rm B} T\spc \frac{m_{\sc 1} x_{\sc 1}+m_{\sc 2} x_{\sc 2}}{m_{\sc 1} m_{\sc 2}}\spc \tau_{\spm \sc w}
=\frac{3}{16}\spc \frac{(k_{\spm\sc\rm B} T)^{2} (m_{\sc 1}+m_{\sc 2})}{p\spc  m_{\sc 1} m_{\sc 2}\O_{\sc 12}^{\sc (1,1)}},
\end{gather}
\begin{gather}
D_{\spm \sc 11}=\frac{3}{8}\spc \frac{(k_{\spm\sc\rm B} T)^{2}}{p\spc  m_{\sc 1} \O_{\sc 11}^{\sc (1,1)}},
\end{gather}
\begin{gather}
D_{\spm \sc 22}=\frac{3}{8}\spc \frac{(k_{\spm\sc\rm B} T)^{2}}{p\spc  m_{\sc 2} \O_{\sc 22}^{\sc (1,1)}},
\end{gather}

\begin{gather}
[\lambda_{\rm th}]_{_{\sc 1}}=\frac{75}{32}\spc \frac{k_{\spm \sc \rm  B}}{m_{\sc 1}}
\spc \dfrac{x_{\sc 1} k_{\spm\sc \rm  B} T}{x_{\sc 1} \O_{\sc 11}^{\sc (2,2)}+x_{\sc 2}\spc \left(\dfrac{25}{2}\spc \O_{\sc 12}^{\sc (1,1)}
-10\spc \O_{\sc 12}^{\sc (1,2)}+2\spc \O_{\sc 12}^{\sc (1,3)}\right)}
+\frac{\rho_{1}[c_{v}^{\rm{int}}]_{_{1}}}{x_{\sc 1}/D_{\sc  11}+x_{\sc 2}/D_{\sc  12}},
\end{gather}

\begin{gather}
[\lambda_{\rm th}^{\sc \rm{tr}}]_{_{\sc 2}}=\frac{5}{2}\spc\frac{k_{\spm \sc \rm B}}{m_{\sc 2}} \spc
 p\spc x_{\sc 2}\spc \tau_{\spm \sc 2}^\prime
=\frac{75}{32}\spc \frac{k_{\spm \sc \rm B}}{m_{\sc 2}}\spc \frac{k_{\spm\sc \rm B} T}{\O_{\sc 22}^{\sc (2,2)}},
\end{gather}

\begin{gather}
[\lambda_{\rm th}^{ \rm{int}}]_{_{\sc 2}}=p\spc [c_{v}^{ \rm{int}}]_{_{2}}x_{\sc 2}\spc \tau_{\spm \sc 2}^{\prime\prime}
=\frac{\rho_{2}[c_{v}^{\rm{int}}]_{_{2}}}{x_{\sc 1}/D_{\sc 12}+x_{\sc 2}/D_{\sc 22}},
\end{gather}
\begin{gather}
k_{\sc T}=\frac{5}{2}\spc x_{\sc 1}x_{\sc 2}\spc \frac{m_{\sc 1} m_{\sc 2}}{\displaystyle (m_{\sc 1} x_{\sc 1}+m_{\sc 2}x_{\sc 2})^{2}}
\dfrac{(5\spc \O_{\sc 12}^{\sc (1,1)}-2\spc \O_{\sc 12}^{\sc (1,2)})}
{x_{\sc 1} \O_{\sc 11}^{\sc (2,2)}+x_{\sc 2} \left(\dfrac{25}{2}\spc \O_{\sc 12}^{\sc (1,1)}
-10\spc \O_{\sc 12}^{\sc (1,2)}+2\spc \O_{\sc 12}^{\sc (1,3)}\right)},
\end{gather}

\begin{equation}
 k_{p} = p\frac{(\partial \mu /\partial p)_{c,T}}{(\partial \mu /\partial c)_{p,T}}= x_{\sc 1} x_{\sc 2} (m_{\sc 2} - m_{\sc 1})\frac{m_{\sc 1} m_{\sc 2}}{(m_{\sc 1} x_{\sc 1} + m_{\sc 2} x_{\sc 2})^3}
\end{equation}

\begin{equation}
\label{dmu/dc}
  \left(\frac{\partial \mu }{\partial c}\right)_{\!\!\spm p,T} = \frac{k_{\spm \sc \rm B}T}{x_{\sc 1} x_{\sc 2}}\frac{(m_{\sc 1} x_{\sc 1} + m_{\sc 2} x_{\sc 2})^3}{(m_{\sc 1} m_{\sc 2})^2}
\end{equation}
where
\begin{gather}
[c_{v}^{ \rm{int}}]_{_{i}}=\frac{3}{2}\spc \frac{(5/3-\gamma_i)}{(\gamma_i-1)}\spc \frac{k_{\spm \sc \rm B}}{m_i}
\end{gather}
is the isochoric specific heat capacity associated with the internal degrees of freedom of molecules of the $i$-component and $\O^{\sc (l,r)}_{\sc ij}$ denotes the elastic Chapman-Cowling collision integrals \cite{Chapman1990}.

Since collision integrals can only be evaluated for a specific interaction between molecules, we shall consider in this paper the Lennard-Jones (6-12) potential function:
\begin{equation}\label{Eq:LJPotential}
  U_{ij}(r) = 4\epsilon_{ij}\left[ \left(\frac{\sigma_{ij}}{r}\right)^{12} -\left(\frac{\sigma_{ij}}{r}\right)^6 \right]
\end{equation}
where $r$ is the distance between the centres of mass of the two molecules, $\epsilon_{ij}$ the maximum depth of the potential well and $\sigma_{ij}$ is the distance at which the potential function vanishes. The values of the Lennard-Jones potential parameters ($\sigma_{ij}$ and $\epsilon_{ij}$) adopted in this model are extracted from literature and listed in Table~\ref{Tab:SphericalParameter}.
\begin{table}
  \centering
  \caption{Parameters for the Lennard-Jones potentials $\sigma_{ij}$ and $\epsilon_{ij}$ for binary gases.}
  \label{Tab:SphericalParameter}
  \begin{threeparttable}
  \begin{tabular}{l l l l l}
  \multicolumn{5}{c}{$\sigma_{ij}$ (nm)}\\
  \hline
        & SF$_6$         &  He             & D$_2$         & H$_2$\\
  SF$_6$& 0.5252\tnote{a}& 0.4298\tnote{a}& 0.4420\tnote{c}& 0.4396\tnote{c}\\
     He &                &  0.2576\tnote{b}&               &        \\
   D$_2$&                &                & 0.2948\tnote{b}&         \\
   H$_2$&                &                &               &  0.2968\tnote{b}\\
\\
  \multicolumn{5}{c}{$\epsilon_{ij}/k_{\sc \rm B}$ (K)}\\
  \hline
       & SF$_6$         & He            & D$_2$         & H$_2$ \\
  SF$_6$& 207.7\tnote{a}&  19.24\tnote{a}& 42.65\tnote{c}&39.77\tnote{c}\\
   He  &                & 10.12\tnote{b}&               &       \\
   D$_2$ &              &               & 39.3\tnote{b} &       \\
   H$_2$&               &               &               &33\tnote{b}  \\
  \hline
\end{tabular}
\begin{tablenotes}
        \footnotesize
        \item[a] Ref.~\cite{Bzowski1990}.
        \item[b] Ref.~\cite{Letamendia1981}.
        \item[c] Calculated based on the Kong rule~\cite{Kong1973}.
    \end{tablenotes}
\end{threeparttable}
\end{table}
The elastic collision integrals $\O^{\sc (l,r)}_{\sc ij}$ are expressed as:
\begin{equation}\label{Eq:ColliIntergal}
  \O^{\sc (l,r)}_{\sc ij} = \sigma^2_{ij}\sqrt{\frac{2\pi k_{\sc\rm B}T}{m_{ij}}}\frac{(r+1)!}{4}\left[ 1 - \frac{1}{2}\frac{\left(1+(-1)^l\right)}{1+l}\right]\O^{\sc \ast(l,r)}_{\sc ij}
\end{equation}
where $m_{i j} = m_{i}m_{j}/(m_{i}+m_{j})$ is the reduced mass, and $\O^{\sc \ast(l,r)}_{\sc ij}$ are the
reduced collision integrals~\cite{Kim2014}, which are functions of the reduced temperature $T^\ast = k_{B}T/\epsilon_{ij}$.

A substantial simplification of our model can be achieved if we consider a mixture of Maxwellian molecules, for which thermal diffusion is automatically
absent.
However, Letamendia and co-workers~\cite{Letamendia1981} have shown that the light scattering spectrum in disparate-mass gas mixtures is very sensitive to the presence and magnitude of thermal-diffusion effects.
An estimate of the contribution of thermal-diffusion effects to the spectral shape was presented by Johnson~\cite{Johnson1983}, who showed that thermal-diffusion effects are comparable in magnitude to other first-order dissipative contributions (heavy-species heat flux and viscosity) in disparate-mass gas mixtures.


Lastly, we close this section by remarking that the calculation of the spectral distribution of scattered light for a disparate-mass gas mixture can be calculated from the dynamic structure factor $S(\boldsymbol q, \omega)$, which was defined in Eq.~\eqref{Eq:StructureMixture}, and further evaluated in the present framework to~Eq.~\eqref{Eq:FinalStructureMixture} under the crucial assumption that $m_{\sc 1}/m_{\sc 2}$ being small. It requires the specification of the molecular masses, polarizabilities, specific heat capacity ratios for all constituents, and Lennard-Jones potential parameters for all combinations of species.
Based on these quantities we can determine: (i) the transport coefficients \eqref{eta-s1} - \eqref{dmu/dc} and (ii) the relaxation times
$\tau_{\sc 2}$, $\tau_{\sc 2}^{\prime}$, $\tau_{\sc 2}^{\prime\prime}$, $\tau_{\spm w}$ and $\tau_{\spm s}$. Thus, the only free adjustable parameter of our generalized hydrodynamic description is the relaxation time $\tau_{\spm \spm \sc v}$, which is connected to the bulk viscosity of the binary mixture via Eq.~\eqref{tau-v}.
Since this is a quantity which cannot be computed within the current framework, we define the so-called internal relaxation number $z=\tau_{\spm \spm \sc v}/\tau_{\spm s}$ representing the ratio between the mean elastic and inelastic molecular collision frequencies.
The value of $z$ can then be determined for each binary mixture from the experimental input. Through the known value of $\tau_{\spm s}=\eta_{\spm s}/p$, the stress relaxation time setting the unit of time, the $z$-parameter is equivalent to:
\begin{equation}
\label{z-par}
    z= \frac{\eta_{\rm b}}{\eta_{\spm s}}  \frac{2}{(\gamma -1)(5-3\gamma)}.
\end{equation}

A similar, but simplified, derivation of the dynamic structure factor $S(\boldsymbol q, \omega)$ can be performed within the framework of classical hydrodynamics, starting from the same matrix equation Eq.~\eqref{Eq:FLtransform}. In this case the matrix elements containing the functional forms $\g(s)$, $f(s)$ and $b(s)$ are then replaced by the expressions in Eqs.~\eqref{g-clas}-\eqref{b-clas}. Rayleigh-Brillouin spectra are computed and a comparison is made with the spectra computed from the complex relaxation hydrodynamics model. A comparison is made for a collisional number of $z=10$ in both cases, but for a variety of scattering angles $\theta$. This value of $z=10$ approximately corresponds to the condition of $p= 1$ bar of SF$_6$ combined with $p= 1$ bar of
helium, for conditions $T$ = 295 K and $\lambda_{i}$= 532 nm (see below).

\begin{figure}
  \centering
  \includegraphics[scale=0.40]{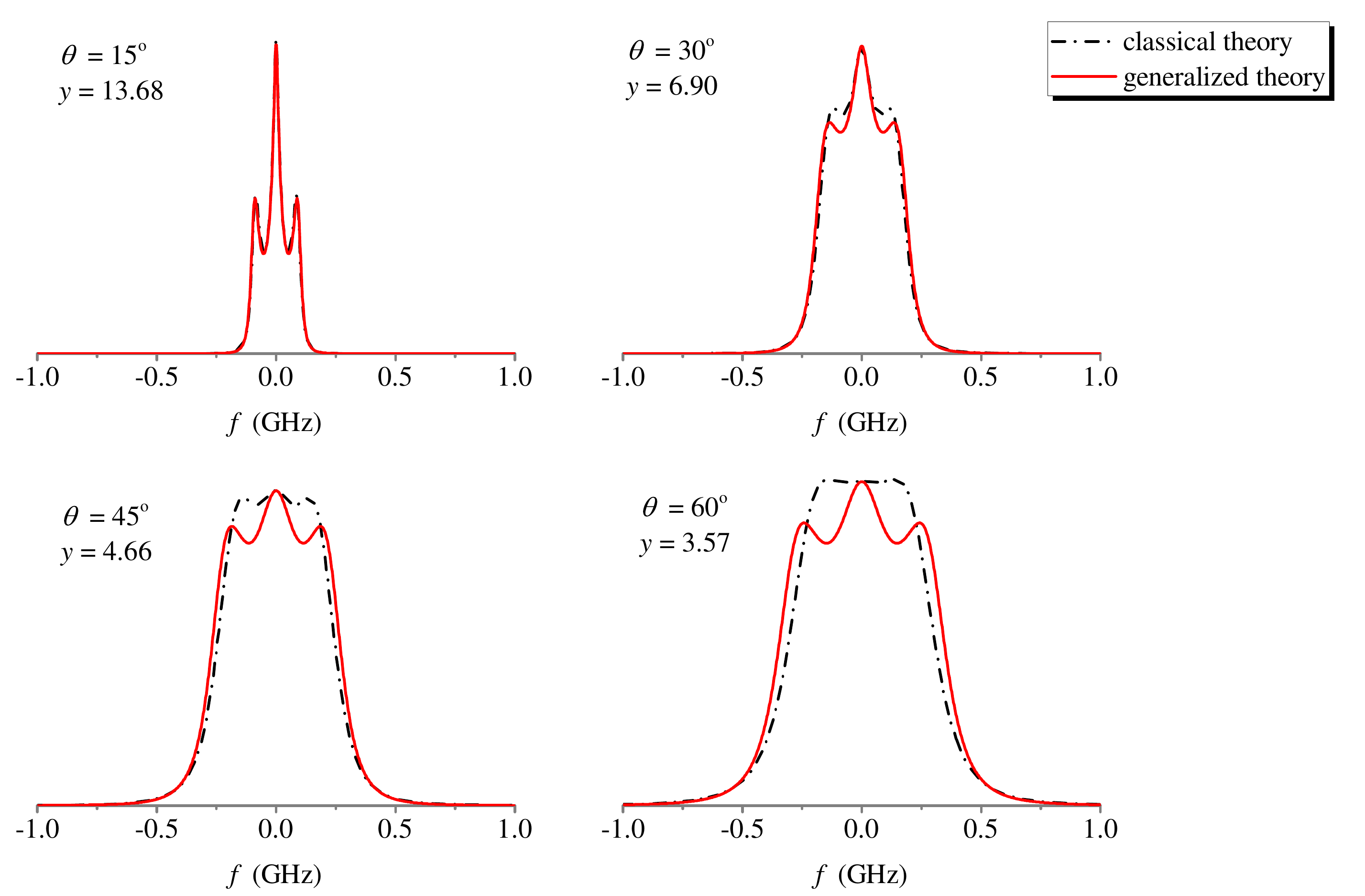}
  \caption{Comparison of computations of the RB-scattering profiles between generalized macroscopic theory vs.  classical hydrodynamics approach for mixtures with internal relaxation number $z=10$ for varying scattering angles $\theta$ and the case of $p=1$ bar SF$_6$ in a mixture with $p=1$ bar of helium.}
  \label{Fig:Comparison}
\end{figure}

The variation of $\theta$ is included in the computations to illustrate the transition from the kinetic to the hydrodynamic regime. The uniformity parameter for the mixture equals:
\begin{equation}
\label{Eq:Parameter-y}
    y= \frac{\lambda_i/n}{4\pi \sin{\theta/2}} \frac{p}{\eta_{\rm s}\sqrt{2k_BT/m}} \ .
\end{equation}
where $p$ is the total pressure and the mass $m$ is to be taken as the mean in the mixture $m=x_1m_1+x_2m_2$.
This representation of $y$ involves an explicit dependence on the scattering angle $\theta$, while all other parameters are kept constant in the comparisons shown in Fig.~\ref{Fig:Comparison}. As indicated, for small scattering angles, hence in the hydrodynamic regime, both theories match closely. Note that $\theta=15^\circ$ corresponds to $y= 13.68$. However, in the kinetic regime of lower values of the uniformity parameter, at $\theta=60^\circ$ corresponding to $y=3.57$, strong deviations are found.

\section{Experimental setup}

\begin{figure}[ht]
  \centering
  \includegraphics[scale=0.40]{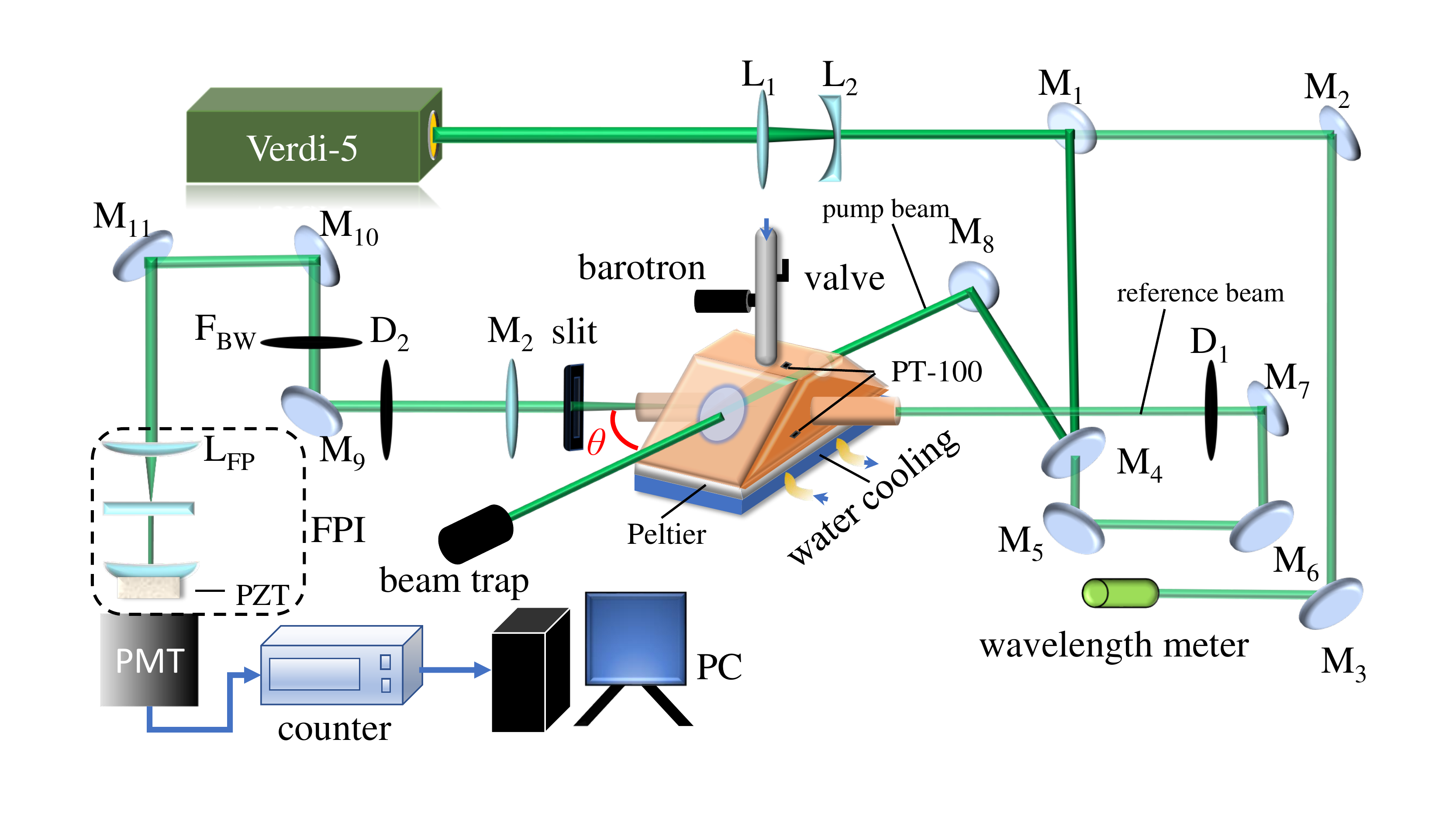}
  \caption{Schematic diagram of the experimental setup for spontaneous Rayleigh-Brillouin scattering at 532 nm.  A Verdi-V5 laser provides continuous wave light at 532.22 nm, at a power of 5 Watt and bandwidth less than 5 MHz. The laser light is split into two beams. The pump beam crosses the RB-scattering gas cell producing scattered light that is captured under an angle $\theta =(55.7 \pm 0.3)^{\circ}$. A small fraction of the power, retained in a reference beam transmitted through M$_4$, is used to align the beam path after the gas cell towards the detector. The scattered light after an bandpass filter (F$_{\rm BW}$) is analyzed in a Fabry-Perot interferometer (FPI), with  free spectral range of 2.9964 GHz and an instrument linewidth of ($58 \pm 3$) MHz, and is collected on a photo-multiplier tube (PMT). Mirrors, lenses and diaphragm pinholes are indicated as M$_i$, L$_i$ and D$_i$. A slit of 500 $\mu$m is inserted to limit the opening angle for collecting scattering light, therewith optimizing the resolution.}
  \label{Fig:GreenSetupSimple}
\end{figure}

The experimental setup used for measuring spontaneous Rayleigh-Brillouin scattering at a wavelength of 532.22 nm, shown in Fig.~\ref{Fig:GreenSetupSimple}, has been described previously~\cite{Wang2019}. The light from a frequency-doubled Nd:VO$_4$ laser (Coherent, Verdi-5), at a power of 5 Watt and bandwidth less than 5 MHz travels through the binary gas medium. For the scattering cell, two Brewster-angled windows are mounted at entrance and exit ports to reduce stray light. A pressure gauge is connected to the cell to monitor the pressure and a temperature control system involving PT-100 sensors. Peltier elements as well as water cooling are used to keep the cell at a constant temperature with uncertainty less than $0.1 \: ^\circ{\rm C}$. The laser wavelength is monitored by a wavelength meter (Toptica HighFinesse WSU-30).

The scattered light propagates through a bandpass filter (Materion, T$>$ 90$\%$ at $\lambda_i$ = 532 nm, bandwidth $\triangle\lambda$ = 2.0 nm) onto a Fabry-Perot interferometer (FPI) with an effective free spectral range (FSR) of $2.9964 (5)$ GHz and an instrument width of $\sigma_{\nu_{\rm instr}}$ = 58.0 $\pm$ 3.0 MHz (FWHM).
The calibration methods were discussed by \citet{Gu2012rsi}. The instrument function is verified to exhibit the functional form of an Airy function, which may be well approximated by a Lorentzian function during data analysis.

For the present experiments a scattering angle $\theta = 55.7 \pm 0.3^\circ$ is adopted, because at angles smaller than the usual setting of $\theta = 90^\circ{\rm C}$ the Brillouin side peaks become more pronounced~\cite{Wang2020,Fabelinskii2012}. The angle was determined by a homemade rotation goniometer stage, while the  opening angle is less than $0.5^\circ$, calculated from the geometry of a slit set behind the gas cell at a certain distance from the scattering center.

RB-scattering spectral profiles were recorded by piezo-scanning the FPI at integration times of 1 s for each step, usually over 18 MHz, with detection of the scattered light on a photomultiplier (PMT) after the FPI analyzer. A full spectrum covering many consecutive RB-peaks and 10,000 data points was obtained in about 3 h. The piezo-voltage scans were linearized and converted to frequency scale by fitting the RB-peak separations to the calibrated FSR-value. The methods for producing such concatenated RB-spectra have been detailed before~\cite{Gu2012rsi,Wang2019PhD}.

\section{Results and comparison}

Rayleigh-Brillouin light scattering spectra were measured for gas mixtures consisting of SF$_6$, the heaviest molecular species to be put in the gas phase at high pressures, combined with gases of the lightest molecules available. Molecular hydrogen (H$_2$) and its isotopologue deuterium (D$_2$) exhibit the same physical and chemical properties, and only the masses are different at 2 amu and 4 amu.
As for a comparison with Helium, its mass is the same as that of D$_2$, while the attractive potential depth for interactions with  SF$_6$ is much smaller. These combinations  provide an interesting test ground for verifying the relaxation hydrodynamic model put forward in section \ref{sec:MixtureModel}. Foremost, the combination of SF$_6$ with low-mass admixed gas fulfills the condition of disparate masses, $m_1<m_2$.
The experimental conditions are chosen keeping a standard pressure of 1 bar SF$_6$, mixed with gases of the lighter species, at pressures stepwise increasing from 0.5 bar to 4 bar.
A list of all pressure combinations experimentally investigated is provided in Table~\ref{Tab:MixtureWithSF6}.
All measurements were performed at room temperature, at $\lambda=532.22$ nm and $\theta=55.7^\circ$.

\begin{table}[hb]
  \centering
  \caption{The conditions of mixture gases experimentally investigated and the fitting result of the internal relaxation number $z$.}\label{Tab:MixtureWithSF6}
  \begin{tabular}{c| c c c||c| c c c || c| c c c}
  \hline
  SF$_6$ &He&&&SF$_6$ &D$_2$ &&&SF$_6$ &H$_2$ &\\
  \cline{1-2}
  \cline{5-6}
  \cline{9-10}
   \multicolumn{2}{c}{$p$ (bar)} & $T$ (K) & $z$ &\multicolumn{2}{c}{$p$ (bar)} & $T$ (K)& $z$& \multicolumn{2}{c}{$p$ (bar)} & $T$ (K) &$z$\\
   \hline
    1.032 &      & 295.0 &  &1.007 &      & 293.2&&1.002 &     & 293.2 &\\
    1.033 & 0.512& 295.1 & 18.73(0.83)&1.001 & 0.506& 293.2& 16.68(0.38)& 1.002 & 0.510& 293.2 &20.75(0.56)\\
    1.037 & 1.029& 295.1 & 9.74(0.36)&1.002 & 1.001& 293.2& 9.91(0.13) &1.002 & 1.004& 293.2 &13.06(0.20)\\
    1.037 & 2.146& 295.1 & 2.44(0.49)&1.002 & 2.003& 293.2& 2.67(0.13) &1.002 & 2.007& 293.2 &5.28(0.06)\\
    1.035 & 2.993& 295.1 & 0.30(0.08)&1.002 & 3.002& 293.2& 1.40(0.07)&1.002 & 3.002& 293.2 &2.51(0.33)\\
    1.037 & 4.084& 295.1 &  $<$ 0.1 &1.002 & 4.002& 293.2& 1.64(0.12)&1.004 & 4.001& 293.2 &1.83(0.12)\\
  \hline
  \end{tabular}
\end{table}

\begin{figure}
  \centering
  \includegraphics[scale=0.5]{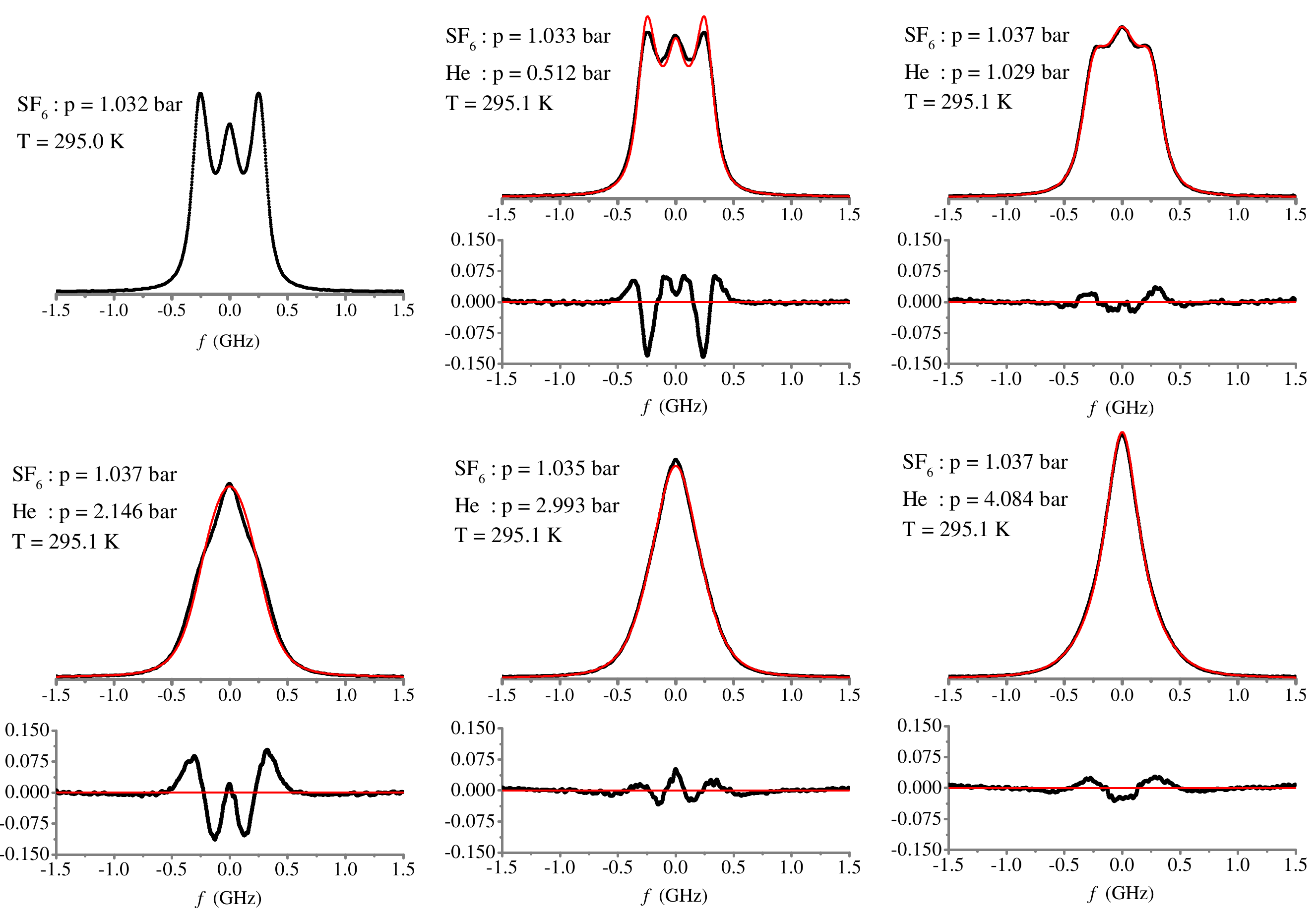}
  \caption{Measured Rayleigh-Brillouin scattering profiles (black) of binary mixtures of SF$_6$-He as measured for the various conditions as indicated, and a comparison with the binary the mixture model (red). Also an RBS-spectrum of pure SF$_6$, at 1 bar, is shown for comparison. Bottom graphs display the corresponding residuals. The experimental data were measured at wavelength of $\lambda_i$ = 532.22 nm and scattering angle of
  $\theta = 55.7^\circ$, and these spectra are on a scale of normalized integrated intensity over one FSR. \label{Fig:SF6andHeExpRes}}
\end{figure}

\begin{figure}
  \centering
  \includegraphics[scale=0.5]{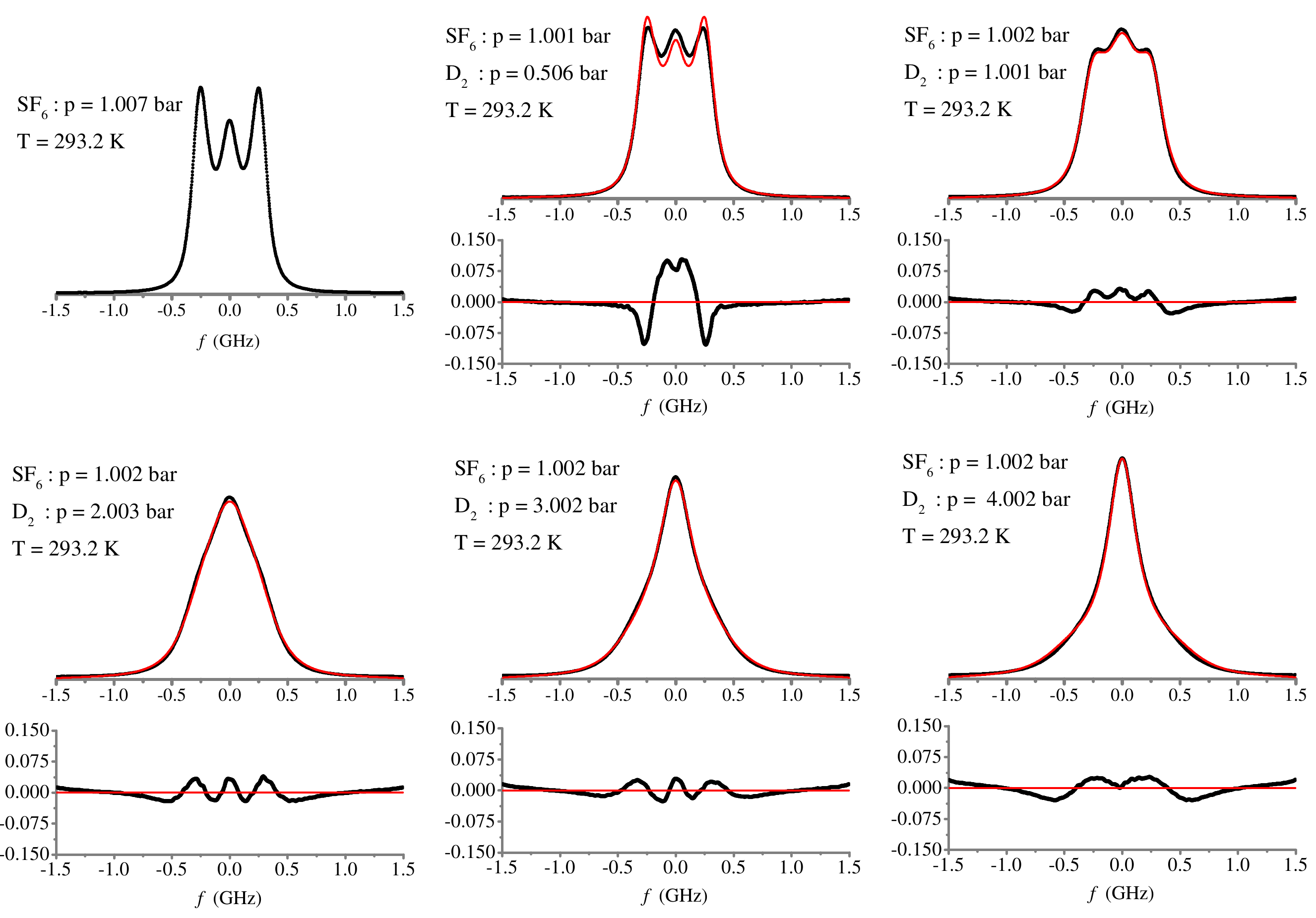}
  \caption{Same as Fig.~\ref{Fig:SF6andHeExpRes} now for SF$_6$/D$_2$ mixtures.}
  \label{Fig:SF6andD2ExpRes}
\end{figure}

\begin{figure}
  \centering
  \includegraphics[scale=0.5]{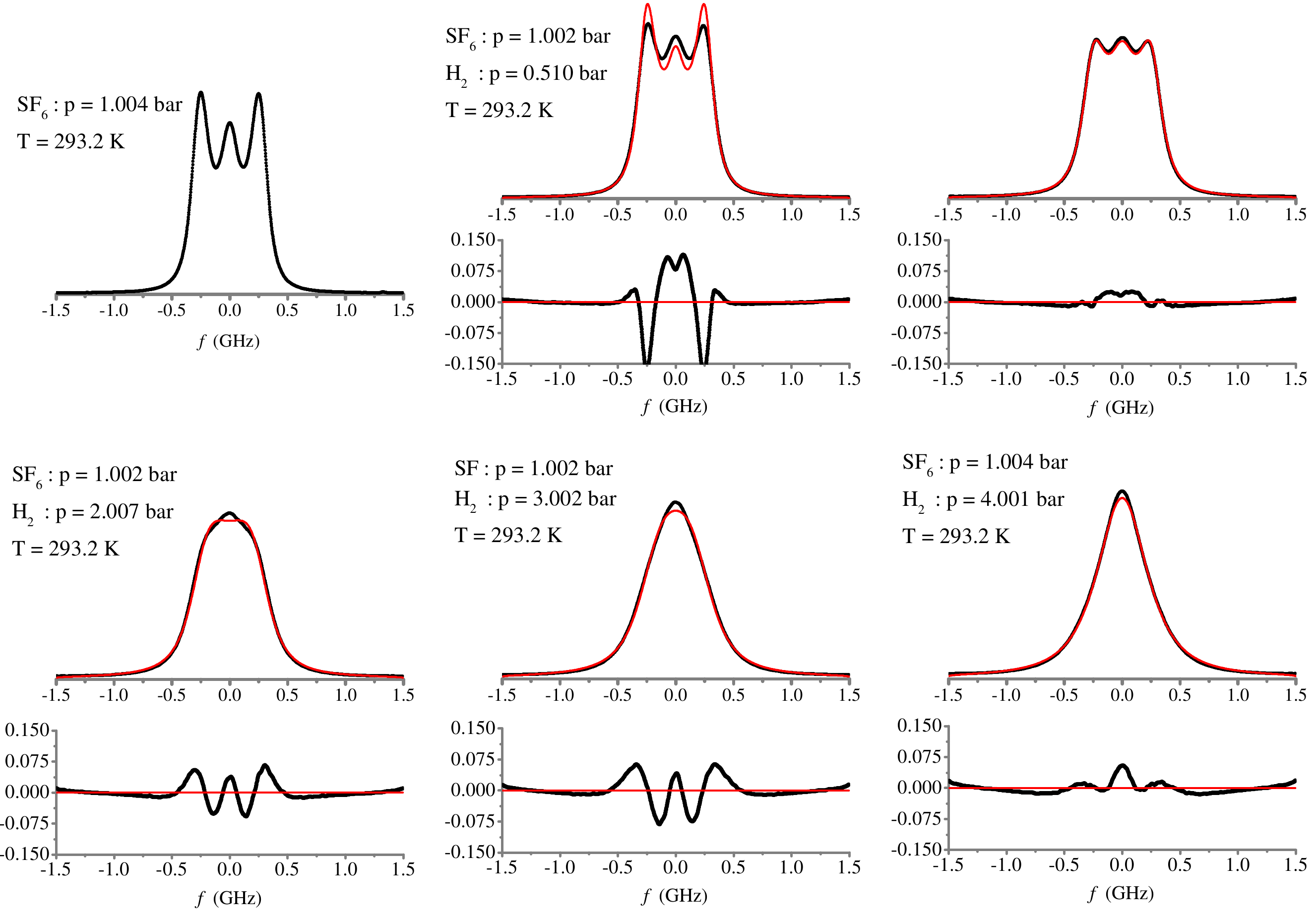}
  \caption{Same as Fig.~\ref{Fig:SF6andHeExpRes} now for SF$_6$/H$_2$ mixtures.
  }\label{Fig:SF6andH2ExpRes}
\end{figure}

The measured RB light scattering spectra, shown in Fig.~\ref{Fig:SF6andHeExpRes} for mixtures with He, in Fig.~\ref{Fig:SF6andD2ExpRes} for mixtures with D$_2$, and in Fig.~\ref{Fig:SF6andH2ExpRes} for mixtures with H$_2$, all show the same qualitative behavior.
In general terms the light scattering of the binary mixtures under investigation is fully dominated by the SF$_6$ molecules.
The polarizability of SF$_6$ is extremely large, causing this molecule to exhibit a very large Rayleigh cross section~\cite{Sneep2005}. In relative terms the polarizability, and therewith the cross section for the light collisional partners is very small, such that these in fact only behave as 'spectators', an effect that was also observed for Ar/He and Kr/He mixtures~\cite{Gu2015}, although not as pronounced as in the present case.



While for single-component gases the Brillouin side peaks become more pronounced when increasing the gas pressure, such as was observed for pure SF$_6$ gas~\cite{Wang2017}, for CO$_2$~\cite{Wang2019}, and for N$_2$O~\cite{Wang2018} in the present case with increasing pressure of the collisional partner in a mixture, the reverse is true. The addition of light-mass constituents to the gas causes the RBS-profile to exhibit less pronounced Brillouin side peaks. A comparison with the profile measured for 1 bar of pure SF$_6$, as shown in the figures, demonstrates this; for the pure SF$_6$ gas the side peaks are most pronounced. A further effect of the addition of light-mass collision partners is the narrowing of the composite RBS profile for increased pressures, and for all three collision partners alike.
So, even though the light collision partners do not contribute to the light scattering themselves, their influence as collision partner is decisive in framing the light scattering spectrum.

The experimentally measured RB-spectra for the various gas mixtures are compared to spectra computed with the theoretical model  for binary gas mixtures as described in section \ref{sec:MixtureModel}. The dynamic structure function $S(\boldsymbol q, \omega)$ is calculated with input from known properties of the SF$_6$ molecule and the light collision partner, which is basically the dynamic polarizability $\alpha$, the heat capacity ration $\gamma$, the coefficients $\epsilon_{i,j}$ and $\sigma_{i,j}$ that define the intermolecular interactions via the Lennard-Jones potential. All gas-transport coefficients needed to evaluate the $S(\boldsymbol q, \omega)$ function are then defined, with the exception of the bulk viscosity of the system, which was parametrized via a single value of $z$ in Eq.~\eqref{z-par}. Extensive computations were performed in which $z$ was treated as a fitting parameter for each individual case of binary mixture in a least squares analysis. The computed spectra for the optimized values of $z$ are plotted in Figs.~\ref{Fig:SF6andHeExpRes} - \ref{Fig:SF6andH2ExpRes}, where also residuals between theory and experiment are presented.
A general trend can be discerned from the comparison between experimental and theoretical spectra. Largest discrepancies occur for the lowest pressure of additions of 0.5 bar of the lighter component, with residuals exhibiting extrema of some 10\% and 15\% for H$_2$. The agreement overall improves for the largest additions of the low-mass scattering partners, where deviations decrease to 2-3\%.

\begin{figure}[hb]
  \centering
  \includegraphics[scale=0.35]{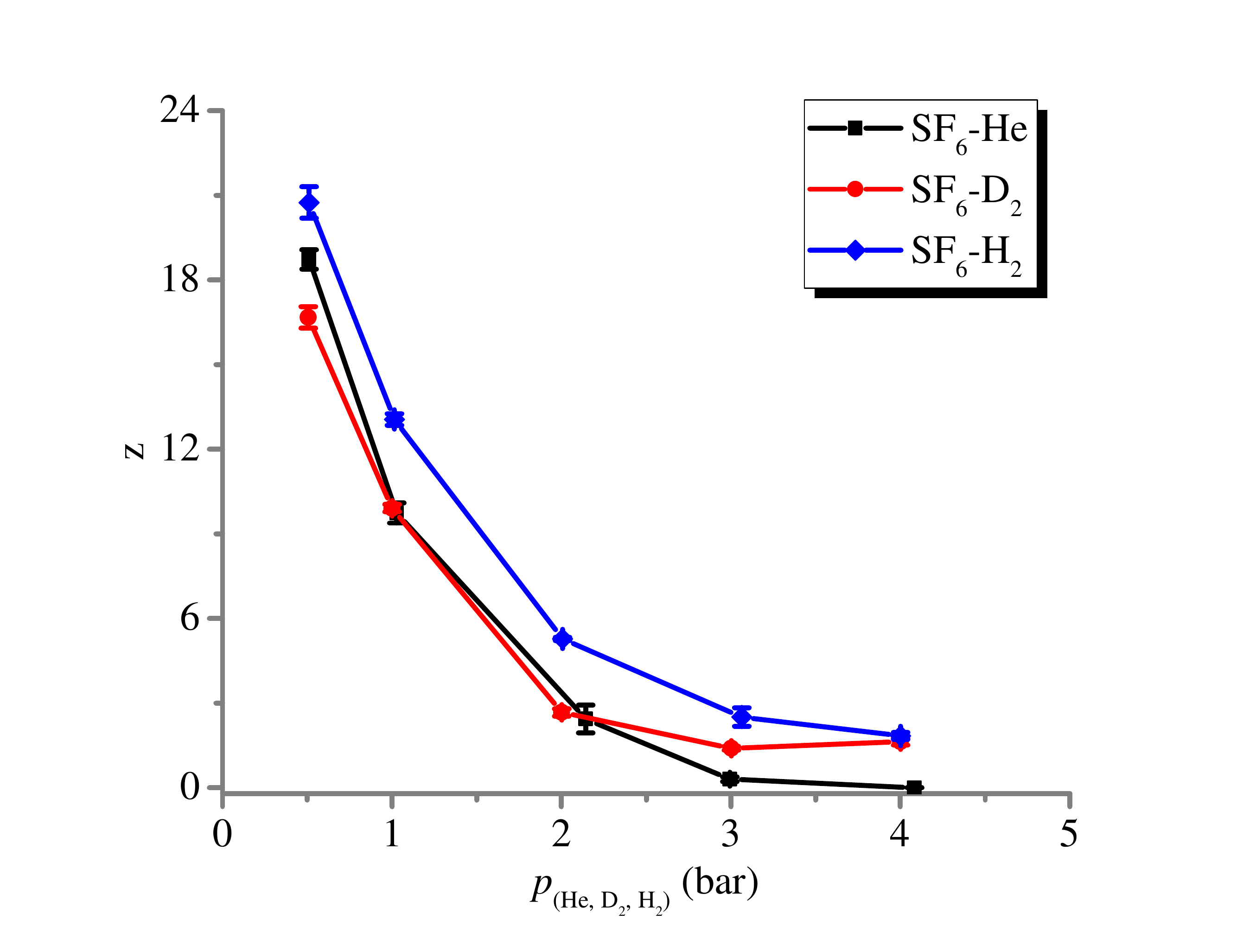}
  \caption{The value of $z$ derived from the comparison between experimental spectra and computed spectra for generalized hydrodynamics, for all pressure combinations and  for the three mixtures of SF$_6-$He, SF$_6-$D$_2$ and SF$_6-$H$_2$. The derived uncertainties are also indicated.
  }\label{Fig:zChangesSF6_3}
\end{figure}

The computations and fitting procedures lead to a set of values for the $z$ parameter, the collisional number for reaching thermal equilibrium between translational and internal energies in the binary mixtures. Results for all combinations of heavy and light species are displayed in Fig.~\ref{Fig:zChangesSF6_3}, while $z$-values and uncertainties are also listed in Table~\ref{Tab:MixtureWithSF6}.
During the fitting optimization of the $z$-parameter it was found that the computed spectra were found to sensitively depend on the values for the heat capacity ratio of $\gamma$(SF$_6$). In these procedures we adopted the experimental value from literature and kept fixed at $\gamma = 1.10$~\cite{Yokomizu2015}. Also the  $\gamma$-values for the light collision partners were set to the literature values as listed in Table~\ref{Tab:PolarizabilityandGamma}. It is noted, that for an optimized value of  $\gamma$(SF$_6$) $= 1.13$, with some differences for various cases of pressures and low-mass mixing partners, a near to perfect match is found between experiment and the generalized RB-scattering theory. This may indicate that the latter is a better value for the the heat capacity ratio, and that the present approach constitutes a manner to determine heat capacity ratios. However, the strong correlations between parameters $z$ and $\gamma$(SF$_6$) in the fitting procedures would require further investigations.

The so-called internal relaxation number $z=\tau_{\spm v}/\tau_{\spm s}$ characterizes the ratio between elastic and inelastic molecular collision frequencies.
As a rule, inelastic collisions - that describe the transfer of energy between translational and internal degrees of freedom - occur less frequently than elastic collisions.
In particular, if the molar fraction of the light constituent increases in a disparate-mass gas mixture, the mean number of inelastic collisions per unit of time suffered by both constituents increases and leads to a decrease of the internal relaxation number.
Based on kinetic gas theory~\cite{Rodbard1990}, it may be verified that in a binary mixture of polyatomic gases the bulk viscosity $\eta_{\rm b}$ of the mixture decreases as the molar fraction of the light constituent increases. Since the parameter $z$ is proportional to the bulk viscosity of the gas mixture, this fact could also explain the decrease of the internal relaxation number as we add light constituents to the mixture.

\begin{figure}
  \centering
  \includegraphics[scale=0.35]{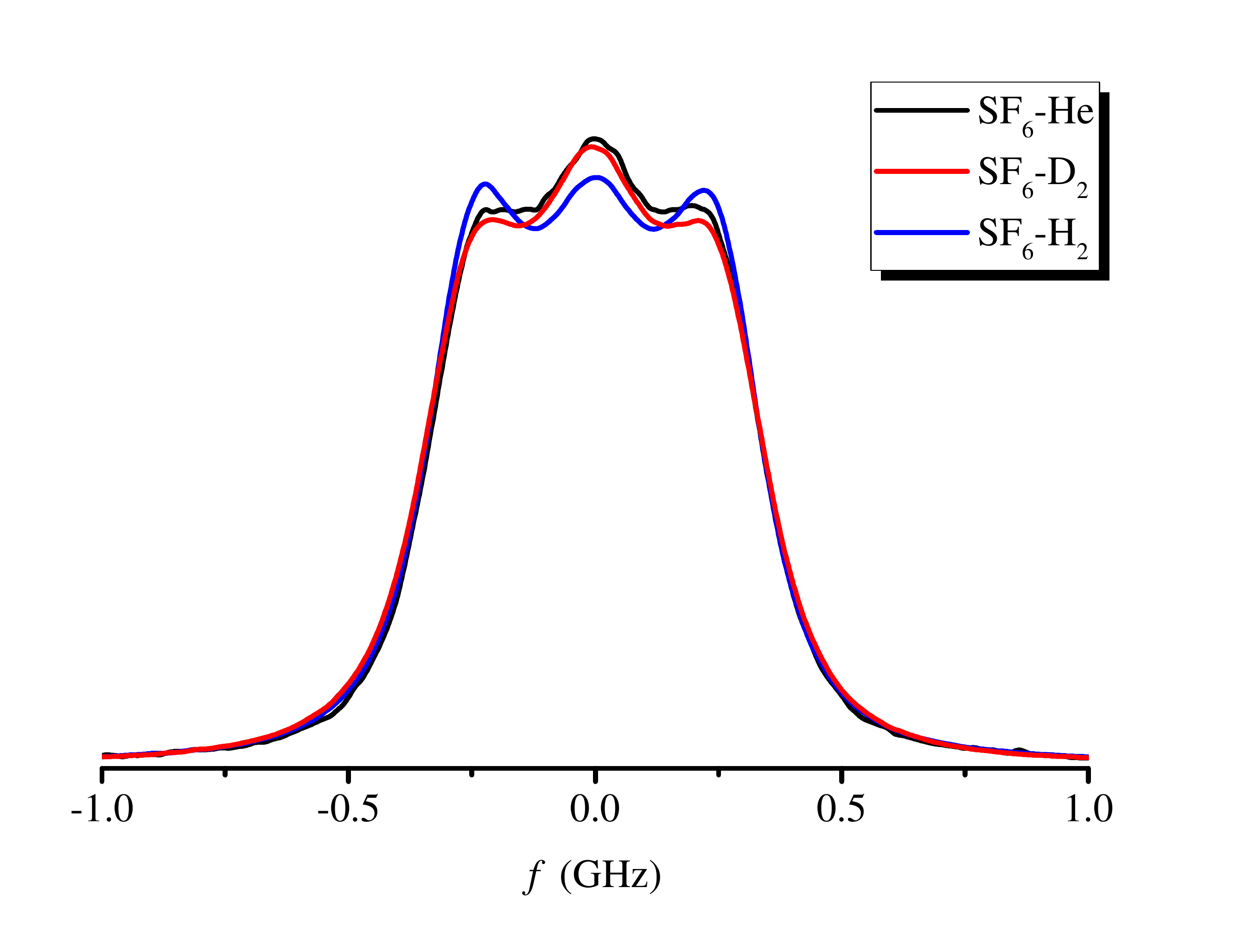}
  \caption{Direct comparison of experimental Rayleigh-Brillouin scattering profiles for 1 bar SF$_6$, admixed with 1 bar of the three light collision partners He, H$_2$ and D$_2$.} \label{Fig:CompareExp-3}
\end{figure}

The trends for the $z$-parameter only show slight differences for the three different collision partners of low mass, but in view of the small uncertainties as resulting from the fitting procedures (cf. Fig.~\ref{Fig:zChangesSF6_3}), these small differences are significant.
In case of H$_2$ as the collision partner the value of $z$ is somewhat larger at a certain pressure, which is indicative for the fact that the heavier species of He and D$_2$ lead to more efficient relaxation.
It should be noted that the generalized hydrodynamic model approach, presented here, makes the assumption that the masses of heavy and light scatterers are strongly disparate, but the mass of the light collision partners does not enter in the model description.
So, in view of the fact that the Lennard-Jones coefficients for collisions between SF$_6$ with H$_2$ or D$_2$ are very similar (cf. Table~\ref{Tab:SphericalParameter}), no difference between H$_2$ and D$_2$ as collision partner would be expected.
While Fig.~\ref{Fig:zChangesSF6_3} compares the resulting values of $z$, in Fig.~\ref{Fig:CompareExp-3} the observed spectra for collision with 1 bar of the light species are compared.
This shows that there is an observable difference for H$_2$ and D$_2$ as collision partners, an effect that goes beyond the current model description.
The fact that the RB-spectra with colliding He and D$_2$ (at 1 bar admixture) are overlapping, leads to the same $z$-parameter for these conditions as shown in Fig.~\ref{Fig:zChangesSF6_3}.
The latter correspondence of $z$-value for He and D$_2$ is indicative of the fact that the specific characteristics of the molecular interaction, as in the $\epsilon_{ij}$ and $\sigma_{ij}$ Lennard-Jones parameters, only plays a marginal role. Only at the highest pressures there arises a deviation between He and D$_2$ as collision partners.

\begin{figure}
  \centering
  \includegraphics[scale=0.35]{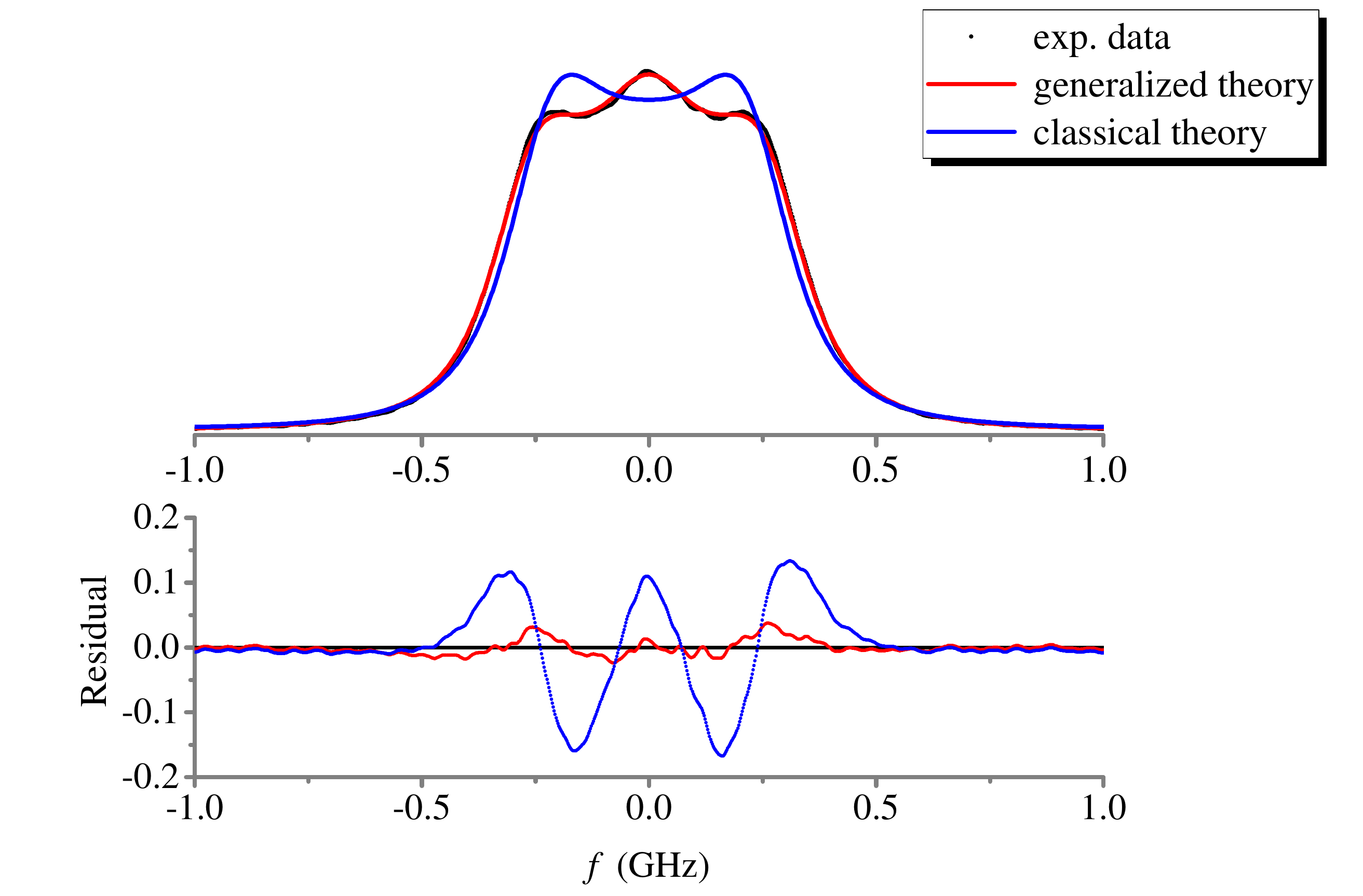}
  \caption{Comparisons between the measured Rayleigh-Brillouin scattering profiles of binary mixtures of SF$_6$-He and generalized relaxation hydrodynamics theory as well as for the classical hydrodynamics approach for mixtures. The experimental data were measured with 1 bar SF$_6$ and 1 bar He at wavelength of $\lambda_i$ = 532.22 nm and scattering angle of
  $\theta = 55.7^\circ$. The theoretical spectra are, for the purpose of comparison with experiment, convolved with the instrumental width of 58 MHz. The spectra are plotted on a scale of normalized integrated intensity over one FSR. \label{Fig:TwoTheorywithExpData}}
\end{figure}

Finally, for an explicit comparison between the observed RBS-profiles and the results from the two theories, generalized relaxation hydrodynamics and classical hydrodynamics, results are plotted in Fig.~\ref{Fig:TwoTheorywithExpData} for the specific case of a 1:1 mixture of SF$_6$/He at $p=2$ bar. This example shows that the generalized theory  gives a superior description of the obtained experimental results, in particular where the kinetic regime is entered in the case of smaller value of $y$.

\section{Conclusion}

A relaxation hydrodynamical model for binary mixture gases is developed that is based on the assumption that the masses of collision partners are disparate. In the model description all macroscopic transport coefficients are computed from the molecular interactions between heavy-heavy, heavy-light and light-light species from a Lennard-Jones potential by invoking the well-tested two-parameter components for the potentials: well depth and characteristic range. Further, the heat capacity ratios $\gamma=c_p/c_v$ and the dynamic polarizabilities $\alpha$ for the gas components are used. From these inputs,  the entire collisional model is produced that allows for a computation of the dynamic structure factor $S(\boldsymbol q, \omega)$ which is representative of the Rayleigh-Brillouin light scattering spectrum.
However, a single ingredient is lacking to complete the calculations: an overall relaxation parameter which can be associated with the bulk viscosity of the binary mixture. This is subsequently set as a fitting parameter in the description of light scattering of binary mixtures.

The model is experimentally tested by performing measurements on binary mixture gases that fulfill the assumption of disparate masses nearly perfectly. For the heavy component, the gas with the heaviest molecular species is chosen that can be brought in the gas phase at high pressure: hexafluoride (SF$_6$). This is combined with the lightest collision partners available, helium gas and hydrogen gas, the latter for two isotopologues H$_2$ and D$_2$. Incidentally the polarizability of the light scattering partners is so small, with respect to that of SF$_6$ that they behave only as spectators in the light scattering process; the light scattered by the light species is negligible for the spectrum.
Nevertheless the activity of the light species as collision partners decisively alters the spectra profiles.
The Brillouin side peaks in the RB-profiles, which are known to become more pronounced at increasing pressure for single species gases, become strongly damped and disappear gradually with higher mole fractions of helium and hydrogen added.

As for a final conclusion the presently developed generalized relaxation hydrodynamics model for Rayleigh-Brillouin scattering in binary gases, based on an assumption of mass-disparate constituents in the gas, provides a good representation of experimentally observed spectral profiles for heavy SF$_6$ in mixtures with light He, D$_2$ and H$_2$. Also the model is shown to be superior to a description in terms of classical hydrodynamics.

\section*{Acknowledgements}
%
YW acknowledges support from the Chinese Scholarship Council (CSC) for his stay at VU Amsterdam.


\end{document}